\documentclass[twocolumn,superscriptaddress,notitlepage,%
aps,longbibliography,prr]{revtex4-2}

\usepackage{amsmath}
\usepackage{amssymb}
\usepackage{mathtools}
\usepackage[svgnames]{xcolor}
\usepackage{graphicx}
\usepackage{adjustbox}
\usepackage{placeins}
\usepackage{csquotes}
\usepackage{tabularx}
\usepackage{blindtext}
\usepackage{comment}
\usepackage{gensymb}
\usepackage{bm}
\usepackage{bbm}
\usepackage{graphicx}
\usepackage{cancel}
\usepackage{ragged2e}

\newcommand{\dd}{\mathrm{d}}
\newcommand{\ee}{\mathrm{e}}
\newcommand{\ii}{\mathrm{i}}

\newcommand{\calE}{\mathcal{E}}

\newcommand{\RR}{{\hat{\mathrm{R}}}}

\newcommand{\hate}{\hat{\mathrm{e}}}

\newcommand{\calL}{\mathcal{L}}

\newcommand{\QED}{\mathrm{QED}}

\expandafter\def\expandafter\normalsize\expandafter{%
    \normalsize%
    \setlength\abovedisplayskip{5pt}%
    \setlength\belowdisplayskip{5pt}%
    \setlength\abovedisplayshortskip{-5pt}%
    \setlength\belowdisplayshortskip{5pt}%
}

\definecolor{light}{gray}{0.90}
\definecolor{darker}{gray}{0.50}
\definecolor{dark}{gray}{0.30}

\definecolor{garrosgreen}{rgb}{0.1, 0.4, 0.1}
\definecolor{dartmouthgreen}{rgb}{0.05, 0.5, 0.06}
\definecolor{duelferred}{rgb}{0.7, 0.2, 0.1}
\definecolor{cambridgeblue}{rgb}{0.1, 0.3, 1.0}
\definecolor{oxfordblue}{rgb}{0.05, 0.2, 0.7}

\begin{document}

\title{Interferometric Differential High--Frequency Lock--In Probe for \\
Laser-Induced Vacuum Birefringence}

\author{R. G. Bullis}
\affiliation{Department of Physics, Colorado State University,
Fort Collins, Colorado 80523, USA}

\author{U. D. Jentschura}
\affiliation{Department of Physics and LAMOR,
Missouri University of Science and Technology,
Rolla, Missouri 65409, USA}

\author{D. C. Yost}
\affiliation{Department of Physics, Colorado State University,
Fort Collins, Colorado 80523, USA}

\begin{abstract}
We propose a measurement of laser-induced vacuum birefringence through the use
of pulsed lasers coupled to femtosecond optical enhancement cavities.  This
measurement technique features cavity-enhanced pump and probe pulses, as well
as an independent control pulse. The control pulse allows for a differential
measurement where the final signal is obtained using 
high-frequency lock-in detection, greatly mitigating
time-dependent cavity birefringence as an important and possibly prohibitive systematic 
effect. In addition, the method features the economical 
use of laser power, and results in a relatively simple experimental setup.
\end{abstract}

\maketitle

\section{Introduction}

In classical electromagnetism, photon-photon scattering is prohibited due to
the linearity of the electromagnetic Lagrangian~\cite{BaRi2013}.
However, in quantum electrodynamics (QED), the presence of virtual
positron-electron pairs introduces nonlinear corrections 
to this Lagrangian density given to
leading order (and in SI mksA units) 
by~\cite{EuKo1935, HeEu1936, Ah2020, HeEu1936, Sc1951, ToFeMiSe2009},
\begin{equation}
\label{heisenberg}
\calL = \frac{\epsilon_0}{2} \left[ \vec E^2 - c^2 \vec B^2 + 
\frac{\alpha}{\pi}
\frac{ (\vec E^2- c^2 \vec B^2)^2+ 7 c^2 (\vec{E} \cdot \vec{B})^2 }%
{45\, E_{\rm cr}^2} 
\right],
\end{equation}
where $\vec{E}$ and $\vec{B}$ are the electric and magnetic fields, 
$\alpha$ is the fine structure constant, 
and $c$ is the speed of light. 
The Heisenberg--Euler Lagrangian~\eqref{heisenberg}
is applicable when the photon energies are much smaller
than the electron rest mass~\cite{HeEu1936}, and the 
amplitudes of the fields are small when compared
to the critical field strengths $E_{\rm cr}$ and $B_{\rm cr}$,
\begin{subequations}
\label{critfields}
\begin{align}
E_{\mathrm{cr}} =& \; \frac{m_e c^2}{|e| \, \cancel{\lambda}_e} 
= 1.3233 \times 10^{18} \, \frac{\mathrm{V}}{\mathrm{m}} \,, \\
B_{\mathrm{cr}} =& \; \frac{E_{\mathrm{cr}}}{c} = 
4.414 \times 10^{9} \, \mathrm{T} \,.
\end{align}
\end{subequations}
These results are obtained by considering the 
energy loss of an electron rest mass energy 
over a reduced Compton wavelength,
i.e., equating $m_e c^2 = |e| \, E_{\mathrm{cr}} \, 
\cancel{\lambda}_e$, where the reduced Compton wavelength 
of the electron is $\cancel{\lambda}_e = \hbar/(m_e c)$ and
$m_e$ is the electron mass.
The first term in Eq.~\eqref{heisenberg}
arises from the Maxwell theory, and the terms proportional to the 
fine-structure constant describe the correction induced 
by virtual electron-positron pairs.

From Eq.~\eqref{heisenberg}, one may derive
nonlinear modifications of the Maxwell equations 
(see, {\em e.g.}, Chap.~18 of Ref.~\cite{JeAd2022book}).
These imply that the quantum vacuum is polarizable; light
traversing a strong electric or magnetic field experiences different indices of refraction
for light polarized parallel and perpendicular to the 
(strong) background field~\cite{KiHe2016, Ah2023, Ah2018,Le2002}. 
This effect is termed vacuum birefringence.  While the
predicted effect is small for typical laboratory magnetic fields, the
direct detection of the vacuum birefringence is an exciting prospect. Besides
validating an as yet unconfirmed prediction of QED, precision tests of the vacuum
birefringence can be used to search for Beyond Standard Model (BSM) physics
since many BSM theories also introduce particles which directly couple to
photons, for example, axions or millicharged particles \cite{AbSi1983, DiFi1983,
Wi1984superstrings, GiJaRi2006prl}. 

A number of experiments aiming to observe vacuum birefringence have been
proposed over the last 60 years.  These include experiments where the
birefringence is established with static magnetic fields \cite{DVEtAl2016,
BeEtAl2012}.  For example, the PVLAS experiment (in Italian: 
{\em Polarizzazione del Vuoto 
con LAser}) uses both superconducting magnets
and permanent magnets to establish a 2.5\,T field, along with high-finesse
optical cavities to measure the induced birefringence.  This experiment has
been ongoing for 25 years and has an uncertainty which is about a factor of 6
larger than the QED predictions \cite{EjEtAl2020}.  

It is also possible to use the magnetic field present in a high-intensity
pulsed laser field to produce a birefringence of the vacuum, which is called
laser-induced vacuum birefringence (LIVB) \cite{Le2002, Ah2023,Ah2018,RoEtAl2021,
SaEtAl2016, Jo1960, Ka2018, ShEtAl2018, HeEtAl2006, LuPe2004, KaEtAl2022, BrEtAl2017,
Ah2024}. Together with laser-induced pair production (see, e.g., the proposals documented in Refs. \cite{BlEtAl2006, ScGiDu2008, KoAhOeSc2022}), the detection of LIVB belongs to the most challenging and most interesting open problems in the physics of strong laser fields. An obvious route to observe LIVB could be based on the use of petawatt laser facilities
\cite{Le2002, Ah2023,Ah2018}. The instantaneous magnetic fields in a petawatt 
laser reach values on the order of $10^6$\,T.  While the fields are very
large, the birefringence is produced over a small region of space (limited by
either the pulse duration or the Rayleigh length of the focus). Therefore, while
LIVB has an advantage due to the large magnetic fields produced, the detection of
vacuum birefringence remains challenging and no demonstration has been
performed to date.

\section{Statistics and Technical Noise}

In typical proposed and 
ongoing measurements, the anticipated vacuum birefringence is
small (on the order of $10^{-10}$\,radians in the PVLAS experiment \cite{EjEtAl2020}).
As a consequence, sufficient statistics requires the detection
of a large number of photons, which is provided by modest optical power.
As a simple example, a 100\,mW visible laser
beam produces about 10$^{17}$\,photons/s. From statistical considerations only 
({\em i.e.}, the standard quantum limit), the ellipticity 
(measured in radians) can be 
determined with a relative uncertainty of
$\approx 1/\sqrt{N_\gamma}$, where $N_\gamma$ is the total number 
of photons measured. Therefore, statistically, 
it should be possible to measure the ellipticity of such a
laser beam with a precision of $10^{-10}$\,radians in about 10 minutes of
integration time. However, technical noise has proven more challenging:
Poorly understood polarization noise associated with the optical cavity mirrors attributed to Brownian and thermoelastic noise of the optical coatings currently 
limits the most recent measurement described in Ref.~\cite{EjEtAl2020}. 
In addition, dynamic mirror birefringence associated with cavity mirror motion 
has been identified as a serious systematic effect in 
Ref.~\cite{HaEtAl2019}. In either case, these 
noise sources are significant at low frequencies 
($\lesssim 1$ kHz), but drop towards higher
frequency.

Therefore, proposed measurements of the vacuum birefringence should first aim
to mitigate technical noise. One of the most powerful ways to accomplish this is 
through lock-in detection, where the system is modulated at a frequency above 
the technical noise sources, and only the system response at that modulation frequency 
is detected, thereby making the system less susceptible to noise. For measurements of
the vacuum birefringence, this is accomplished through the modulation of a
strong background field, which can be 
electric or magnetic, or a combination thereof.
For measurements using a quasi-static magnetic
field, this modulation can be accomplished through pulsing~\cite{BeEtAl2012}, or
rotating the magnetic field~\cite{GiJaRi2006prl}.  In the case of LIVB, the signal is
naturally modulated at the pulse repetition rate of the high-power laser, which
could be used for lock-in detection.  For all these examples, it is difficult
to obtain lock-in frequencies greatly above $\sim$10\,Hz, which
is somewhat dissatisfying because
there is a strong motivation to increase the modulation 
frequency well above the frequency
of prevalent lab noise sources.

%
%
\section{Interferometric Lock--In Method}

\begin{figure}[t!]
\centering
\includegraphics[scale=0.45]{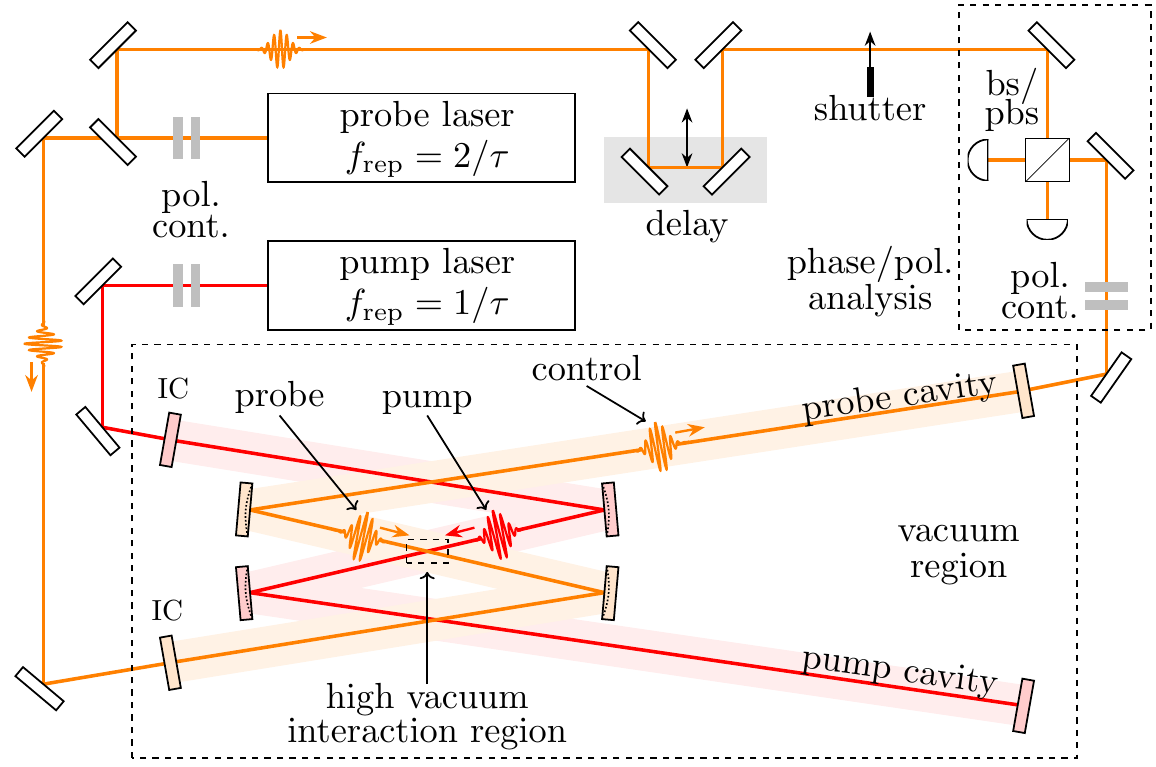}
\caption{In the proposed experimental
setup, two modelocked laser systems (laser oscillators followed by amplification)
are coupled to pump and probe FECs
through two input couplers (IC), which would have transmissions of 
$\sim 0.2$ \% for the 
pump cavity and $\sim 0.04$ \% for the probe cavity. 
Parabolic mirrors (5 cm focal length)
produce a tight intracavity focus for both FECs. The probe pulse interacts with the
pump pulse and acquires an ellipticity, 
$\chi_{\text{QED}}$, 
within a high-vacuum region which is differentially pumped and maintained 
within a lower vacuum region containing the FECs. 
The high vacuum region is necessary is to prevent systematic effects
related to residual gas in the interaction region \cite{BuPrWh1967}. 
Since the probe laser repetition rate is 
twice that of the pump laser, a control pulse will also exist within the probe cavity
but is timed to never interact with the pump pulse at the focus. Cavity transmission
is used for RF polarimetry and lock-in detection at $\sim$100 MHz.
Alternatively, an interferometric measurement of $n_1$ and $n_2$ can be made
separately by adjusting the input polarization, replacing the polarizing
beamsplitter (pbs) with a non-polarizing one (bs), and interfering the cavity transmission
with the incident pulse train. Depending on which measurement is made, the polarization
will need to be adjusted with waveplates (pol. cont.)}
\label{fig1} 
\end{figure}

In order to increase the lock-in frequency, it is interesting to consider a method to
observe the vacuum birefringence, 
following, to a certain extent, ideas presented in Ref.~\cite{LuPe2004}.
We aim to utilize optical cavities to provide energy enhancement of femtosecond optical
pulses.  These cavities are termed femtosecond enhancement
cavities (FECs) and have mostly been applied to intracavity high-harmonic
generation and extreme ultraviolet frequency 
combs~\cite{JoMoThYe2005,PeEtAl2005,MoJJYe2005}, and have also been 
used to suppress noise and increase sensitivity 
in transient absorption spectroscopy \cite{ReChAl2015}.
Here, we propose (see Fig.~\ref{fig1}) to use
two FECs, each comprising at least four mirrors, two of which are curved,
in order to individually power-enhance two optical
pulses, a pump and a probe, which are made to repeatedly collide at a focus
(the interaction region)
shared between the two optical cavities \cite{LuPe2004}.
The pulse energy of the pump pulse is assumed to be much smaller 
than currently available at petawatt laser facilities, but could still 
reach mJ pulse energies with $\sim$250~fs durations.  When focused 
to a Gaussian beam waist radius of $\sim$4 $\mu$m, this results in peak magnetic
fields of $\sim$10$^3$ T. One would
compensate for the relative deficiency in the peak fields by a setup 
where the probe pulse samples the birefringence repeatedly, producing a cumulative polarization rotation. This method uses the laser pulse energy
economically, since it is continually recycled through the use of the optical
cavities~\cite{JoMoThYe2005}.

For reference, we point out that, while we use linear geometries for the
cavities (as opposed to ring geometries), there are no standing waves within
the FECs, but the laser pulses propagate inside the FECs in a ``ping-pong''
fashion, where the timing is adjusted so that the probe and pump pulses
repeatedly overlap in the focus.  More importantly, the use of FECs offers new
possibilities to increase the lock-in frequency of the measurement drastically
( {\em i.e.}, the rate at which the pump pulse periodically enters the
interaction region).  Within an FEC, this rate is typically on the order of
about~$100$\,MHz~\cite{JoMoThYe2005}.  An enabling feature of our proposal is
the addition of a second ``control'' pulse to the probe enhancement cavity.
This control pulse is timed so that it does not interact with the pump pulse.
Differential measurements between the probe and control pulses effectively
enable the high-frequency lock-in measurement.

The basic experimental idea is shown in Fig.~\ref{fig1} and starts with two modelocked
lasers. As is well known, a standard modelocked oscillator produces a single short
pulse oscillating between the laser cavity mirrors. 
We denote by $\tau$ 
the time it takes for the pump pulse to make one round trip in the laser cavity.
This single-pump laser pulse
is repeatedly sampled at the output coupler and generates a pulse train
external to the laser cavity with a 
repetition rate given by $f_{\rm rep}=1/\tau$ \cite{Cu2002}.
By coupling this pulse train to an enhancement cavity with a round-trip
time $\tau_{\rm enh}=\tau$, the pulse train can be coherently stacked, producing a
single pulse again in the enhancement cavity~\cite{JoYe2002}. 
As a technical aside, the cavity lengths
will be locked to the pump and probe seed lasers using the 
Pound--Drever--Hall method \cite{PDH}.
In addition, the offset frequency of the pump and 
probe seed lasers will be adjusted to precisely
match the round-trip phase evolution of the intracavity pulses arising from 
cavity dispersion~\cite{JoYe2002}.
While the pulse energy can be greatly enhanced by the cavity, 
the establishment of a high-energy ($>1$\,mJ) intracavity 
pulse is greatly assisted by high-power fiber amplification 
of the incident pulse train~\cite{LiEtAl2016}.

The high-energy pump pulse establishes the vacuum birefringence at a tight
intracavity focus.  Here, we envision the focus being formed by two parabolic
mirrors with focal lengths of around 5~cm.  The birefringence at the focus is
probed with a similar pulse established in a separate enhancement cavity which
also has a round trip time, $\tau$.  However, the probe enhancement cavity will
be excited by a modelocked laser with a repetition rate of $f_{\rm
rep}=2/\tau$.  With this, two pulses, a probe and a control pulse,
simultaneously exist within the probe cavity.  While this requires the precise
synchronization between the repetition rates of the probe and pump lasers, the
feasibility of such synchronization has been demonstrated at a level which far
exceeds the requirement of this proposal using standard phase-locking
techniques~\cite{MaEtAl2001}.

Using this method for the generation of the probe and control pulses, it is
clear why they will be nearly identical in terms of polarization, pulse energy,
pulse duration, and phase evolution (the only difference is a relative time
delay within the probe cavity).  Namely, the ultimate source of both pulses is
the same pulse propagating within the probe laser cavity~\cite{Cu2002}.  In
addition, the beam path to the probe enhancement cavity and the optical path
within the probe enhancement cavity are necessarily identical for both the
probe and control pulses.  Because of the different repetition frequencies of
the probe and pump lasers, the control pulse never overlaps with the pump
pulse.  The differential polarization ellipticity between the probe and control
pulses established by the vacuum birefringence can be measured using the
leakage light from the probe cavity.

\section{Theory}

\subsection{QED Predicted Birefringence}

\begin{figure}[t!]
\centering
\includegraphics[scale=0.5]{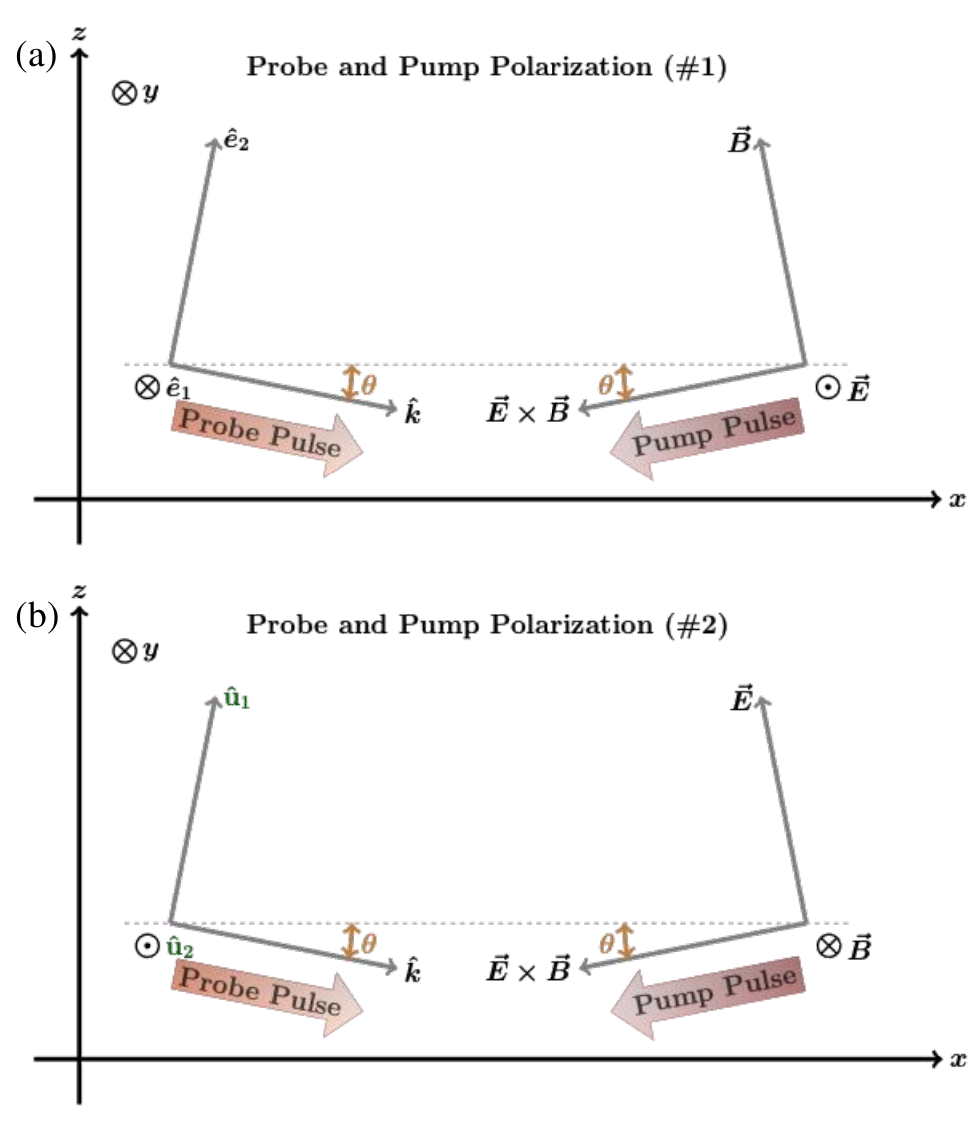}
\caption{Diagram showing the relative orientation of pump and probe polarization and propagation direction for two potential measurement scenarios. (a) The orientation or the pump and probe polarizations considered in the text. (b) Alternative pump and probe polarizations.}
\label{fig2} 
\end{figure}

Here, we consider two relative orientations of the pump and probe polarizations and propagation directions as shown in Fig.~\ref{fig2}. First, assuming that the pump pulse polarization is out of the page, the vacuum birefringence is just a phase shift between the two  polarizations of the probe pulse, which are horizontal and (nearly) vertical as shown in Fig.~\ref{fig2}a.
In order to determine the signal size, 
the effective refractive indices relevant for the
probe pulse colliding with an intense probe pulse at a slight angle, $\theta$,
are required. Given $\hat k$ as the propagation direction of the probe field
at the interaction region in Fig.~\ref{fig1},
and $\vec E$ and $\vec B$ denoting the electric and magnetic fields of the pump
pulse, we define 
the polarizations of the probe field in Fig.~\ref{fig2} as follows, 
\begin{align}
\vec e_1 =& \; \vec E - (\hat k \cdot \vec E) \, \hat k 
+ c\, \hat k \times \vec B \,,
 \qquad
\hat e_1 = \frac{\vec e_1}{|\vec e_1|} \,,
\\
\vec e_2 =& \; c \,\vec B  - 
c \, ( \hat k \cdot \vec B )  \, \hat k - \hat k \times \vec E \, , \qquad
\hat e_2 = \frac{\vec e_2}{|\vec e_2|} \,.
\end{align}
The two polarizations of the probe,
in the coordinate system given in Fig. 2a,
have the representation
\begin{subequations}
\label{pol1}
\begin{align}
\hat e_1 =& \; \hate_y \,,
\\
\hat e_2 =& \; \hate_x \, \sin\theta + \hate_z \, \cos\theta \,,
\qquad \hat e_1 \times \hat e_2 = \hat k \,.
\end{align}
\end{subequations}

We now consider the 
QED contributions $\Delta n_1 = n_1 - 1$ and $\Delta n_2 = n_2 - 1$ 
to the refractive indices corresponding to the polarizations 
$\hat e_1$ and $\hat e_2$.
They can be obtained from Eq.~\eqref{heisenberg} and are given 
by~\cite{Baier1967, Na1969qed,Ka2016,DiGi2000}
\begin{subequations}
\label{n1n2gen_pol1}
\begin{align}
\Delta n_1 =& \frac{4 \alpha}{90\pi}
\frac{(\hat k \! \times \! \vec E)^2 + 
c^2 \, (\hat k \! \times \! \vec B)^2 -
2 c \, \hat k \cdot (\vec E \! \times \! \vec B)}{E_{\mathrm{cr}}^2 }  \,,
\\
\Delta n_2 =& \frac{7 \alpha}{90\pi}
\frac{(\hat k \! \times \! \vec E)^2 + 
c^2 \, (\hat k \! \times \! \vec B)^2 -
2 c \, \hat k \cdot (\vec E \! \times \! \vec B)}{E_{\mathrm{cr}}^2 }  \,.
\end{align}
\end{subequations}
Here, $\vec E$ and $\vec B$ are the instantaneous 
electric and magnetic fields of the pump pulse. 
According to Eq.~\eqref{n1n2gen_pol1},
for co-propagating probe and pump fields, the 
induced birefringence vanishes, while $\Delta n_1$ and $\Delta n_2$
are enhanced for the near-counter-propagating setup
of Fig.~\ref{fig2}.

With the geometry for the proposed experimental 
setup shown in Fig.~\ref{fig2}a
(with the pump pulse polarization along the $y$ axis), the QED contributions to the indices of refraction and to the birefringence evaluate to 
\begin{subequations}
\label{n1n2exp}
\begin{align}
\label{n1exp}
\Delta n_1 =& \; n_1 - 1 = \frac{4 \alpha}{90\pi}
\frac{4 E^2 \cos^4 \theta}{E^2_{\mathrm{cr}}} \,,
\\
\label{n2exp}
\Delta n_2 =& \; n_2 -1 = \frac{7 \alpha}{90\pi}
\frac{4 E^2 \cos^4 \theta}{E^2_{\mathrm{cr}}} \,,
\\
\label{nQED}
\Delta n_{\QED} =& \; n_2 - n_1 = \frac{\alpha}{30\pi}
\frac{4 E^2 \cos^4 \theta}{E^2_{\mathrm{cr}}} \,.
\end{align}
\end{subequations}
For an experimentally reasonable $\theta=10^\circ$, the signal 
will decrease minimally
(94\% of the maximum). We note that these indices depend on the magnitude 
squared of the fields and are, therefore, 
not sensitive to the relative optical phase of the
pump and probe pulses.

Next, we consider the geometry in 
Fig.~\ref{fig2}b with the axes of the electric and magnetic  fields of the pump laser 
(and of the probe laser) interchanged relative to Fig.~\ref{fig2}a.
Still, we denote by $\vec E$ and $\vec B$ the 
instantaneous electric and magnetic fields
of the pump laser. Then, 
we define the polarizations $\hat u_1$
and $\hat u_2$ of the probe laser in the same way as Eqs. (3) and (4).
The two polarizations of the probe,
in the coordinate system given in Fig.~\ref{fig2}b,
are [cf.~Eq.~\eqref{pol1}]
\begin{subequations}
\label{pol2}
\begin{align}
\hat u_1 =& \; \hat u_x \, \sin\theta + \hat u_z \, \cos\theta \,,
\\
\hat u_2 =& \; -\hat u_y \,,
\qquad \hat u_1 \times \hat u_2 = \hat k \,.
\end{align}
\end{subequations}

The general formulas for the 
refractive indices $n_1$ and $n_2$
which belong to the polarizations
$\hat u_1$ and $\hat u_2$, respectively,
given in Refs.~\cite{Na1969qed,Ka2016,DiGi2000},
still remain valid.
However, we should remember that these
formulas now pertain to the 
polarizations $\hat u_1$ and $\hat u_2$
as given in Fig.~\ref{fig2}b,
which are interchanged with respect to the 
polarizations used in Fig.~\ref{fig2}a,
in view of the fact that the $\vec E$ 
and $\vec B$ fields also are interchanged. These formulas are formally equivalent to those
given in Eq.~\eqref{n1n2exp},
but, again, with the physical interpretation of 
the polarizations interchanged such that 

\begin{equation}
\label{nQED2}
\Delta n_{\QED} = \; n_1 - n_2 = -\frac{\alpha}{30\pi}
\frac{4 E^2 \cos^4 \theta}{E^2_{\mathrm{cr}}} \,.
\end{equation}

Hence, for a differential measurement
of $\Delta n_{\QED}$, we can 
expect a sign reversal of the experimental
signal under a flip of the pump laser 
polarization.

\begin{figure}[t!]
\centering
\includegraphics[width=0.93\linewidth]{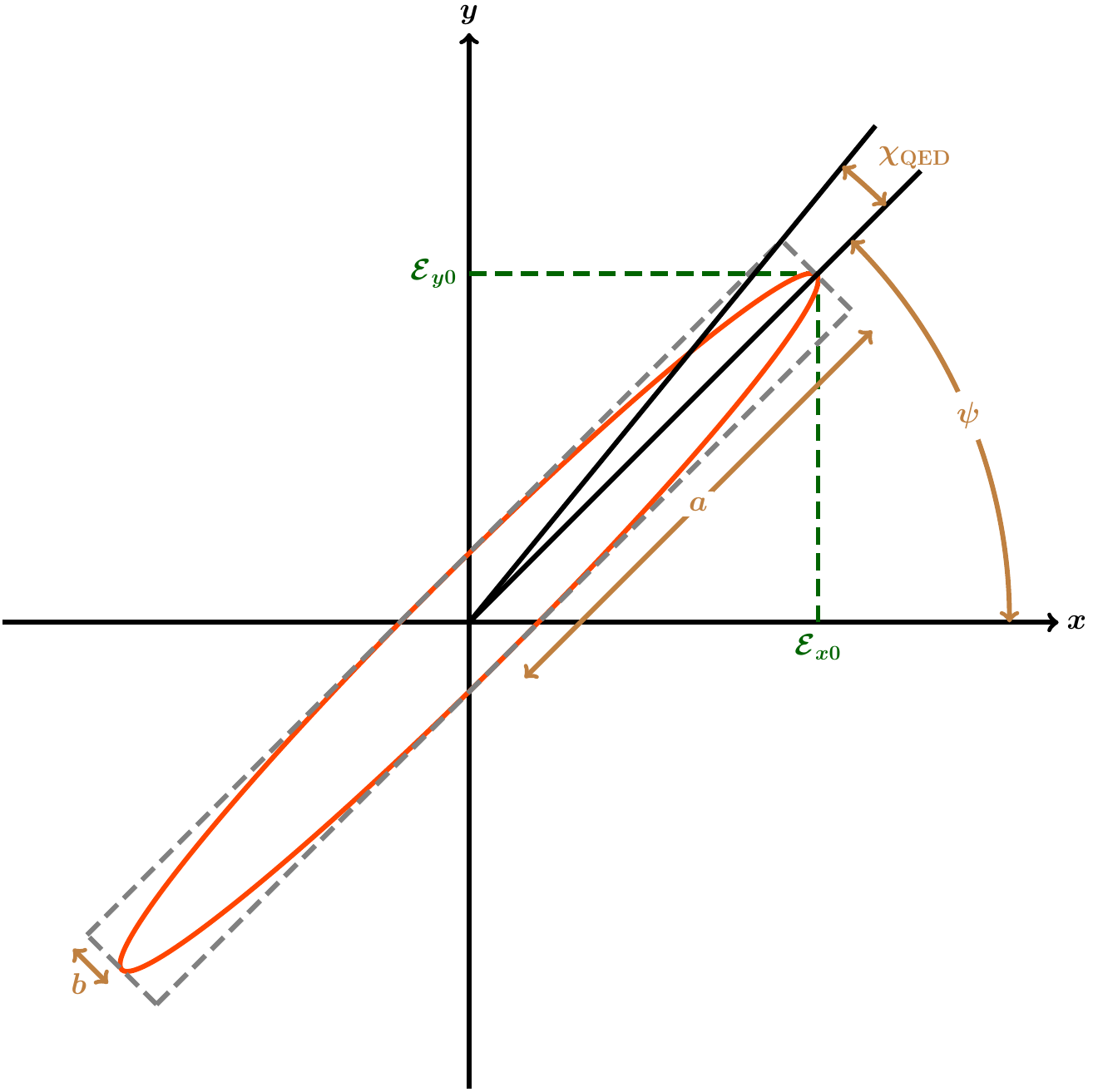}
\caption{We consider the geometry of
the polarization ellipse generated 
for an oncoming pulse of orientation $\Psi$,
induced by vacuum birefringence.
The eccentricity angle is $\chi_{\QED}$.}
\label{fig4}
\end{figure}

\subsection{Polarization Ellipse}

We aim to clarify the
relation of the orientation angle $\Psi$
of the probe laser,
and of the eccentricity angle $\chi_\QED$
of the polarization ellipse of the 
probe after passing through the interaction
region. For definiteness, let $z$ be the propagation direction of the 
probe. (For reference, one notes that this
orientation of the $z$ axis is different 
from the one used in Fig.~\ref{fig2})
The definitions of the orientation angle $\Psi$,
and of the eccentricity angle $\chi_\QED$,
of the polarization ellipse,
of the probe laser, after
passing through the field of the pump laser,
are as indicated in Fig.~\ref{fig4}.
The phase difference of $x$ and $y$ components of
the electric field is denoted as $\delta$.
Let $\calE_x(z,t)$ and $\calE_y(z,t)$ now denote
the electric-field components of the probe laser, 
notably, directed along axes 
perpendicular to its propagation direction,
which is the $z$ axis.
The equation of the polarization ellipse
is found to be~\cite{Co2005}
\begin{equation}
\frac{\calE_x(z,t)^2}{\calE_{0x}^2} +
\frac{\calE_y(z,t)^2}{\calE_{0y}^2} -
\frac{2 \calE_x(z,t) \calE_y(z,t)}{\calE_{0x} \, \calE_{0y}} \,
\cos \delta = \sin^2 \delta \,.
\end{equation}
The orientation angle $\Psi$ and the eccentricity
angle $\chi$ are found as follows,
\begin{subequations}
\begin{align}
\tan(2 \psi) =& \; \frac{2 \calE_{0x} \, 
\calE_{0y}}{\calE_{0x}^2 - \calE_{0y}^2} \,
\cos\delta \,,
\qquad 0 \leq \Psi \leq \pi \,,
\\
\sin(2 \chi) =& \; \frac{2 \calE_{0x} \, 
\calE_{0y}}{\calE_{0x}^2 + \calE_{0y}^2} \sin\delta \,
\qquad -\pi/4 \leq \chi \leq \pi/4 \,.
\end{align}
\end{subequations}
One notes that 
$\psi = \pi/4$ for 
$\calE_{0x} = \calE_{0y}$.
Furthermore, expressed in terms of the 
axes of polarization ellipse which
has a semi-major axis of length $a$,
and a semi-minor axis of length $b$,
the eccentricity parameter is found as
$\chi = \arctan(a/b)$.
For $\calE_{0x} = \calE_{0y}$, one finds that
\begin{equation}
\label{chiQED}
\chi_\QED = \frac12 \delta
= \frac12 \, N \, \left( \frac{2 \pi L}{\lambda} \right) \, \Delta n_{\QED} 
= \frac{\pi N L}{\lambda} \, \Delta n_{\QED} \,,
\end{equation} 
where $N$ is the number of traverses through
the interaction region, $L$ is the 
interaction length, $\lambda$ is the 
probe laser wavelength, 
and $\Delta n_{\QED}$ is the differential
refractive index in the interaction 
region induced by vacuum birefringence
in the field of the pump laser~\cite{Le2002}.
\vspace{.5 cm}

\subsection{Expected Ellipticity}

In order to benchmark the expected signal, 
we first focus on the expected ellipticity
$\chi_{\QED}$ of a probe pulse whose polarization axis 
is at a $45^\circ$ angle between 
the $\hat{e}_1$ and $\hat{e}_2$ directions. This is given by Eq.~\eqref{chiQED}.
In practice, $L$ will be limited by either
the Rayleigh length or the effective length of the 
temporal probe and pump pulses (whichever is smaller)~\cite{Le2002}. 

To estimate $\chi_{\QED}$, we 
 assume  state-of-the-art
FEC performance, as described in Ref.~\cite{CaEtAl2014megawatt},
namely, an intracavity pulse
energy of~1.6\,mJ with a pulse duration of 250\,fs at a laser wavelength
of $\lambda=1040$\,nm (photon energy of 1.19\,eV). A key result of that work 
was the self-limiting thermal behavior of the cavities allowing for 
an intracavity power of the order of $10^6$\,W.
As an aside, we note that, 
while a petawatt laser has a peak power which is about 5 orders of magnitude larger 
that the FEC system demonstrated in~\cite{CaEtAl2014megawatt}, the 
FEC system average power can exceed that of a petawatt laser 
by about three orders of 
magnitude due to the much higher effective repetition rate.
If one produces a Gaussian beam waist radius of
4~$\mu$m at the focus, which is chosen so that the Rayleigh 
length (given by $z_\RR=\pi w_0^2/\lambda$, 
where $w_0$ is the Gaussian radial width at the focus)
roughly matches the pulse
duration, then, at the intracavity focus, the peak electric field of the pump
pulse is $|\vec{E}| \approx \, \, 4 \times \,10^{11}$\,V/m, and the peak 
magnetic field is $|\vec{B}| \approx \, \, 1.5 \times \,10^{3}$\,T. This results
in a peak $\Delta n_{\QED}$ of $\approx 3 \, \times 10^{-17}$, 
which must be divided by two in order to account for
the oscillatory nature of the pump pulse. 

We envision that the high-power
FEC of the pump laser would have a finesse of $\sim$5,000 \cite{CaEtAl2014megawatt} 
while the probe cavity would have a finesse of $\sim$30,000.  This asymmetry 
in cavity finesse is chosen because, for the pump cavity, only the intracavity power
(and not the power buildup) is relevant.
However, for the probe cavity, the buildup directly multiplies
the signal and so it is advantageous to have a higher finesse cavity. 
With this, the probe enhancement cavity power
buildup can realistically exceed $N=10^4$, so that 
$\chi_{\QED} \, \approx \,
1.9 \times 10^{-11}$. As discussed in Appendix B, when taking into account a Gaussian spatiotemporal
pulse shape of both the probe and pump pulse, this peak ellipticity is modestly reduced
to $\chi_{\QED} \, \approx \, 0.8 \times 10^{-11}$.
As discussed above, from a statistical perspective, the
detection of this small ellipticity appears to be perfectly 
feasible.  With $\sim 1$\,W of cavity
output light available for the measurement 
($\sim 5 \, \times\, 10^{18}$ photons/s), $\chi_{\QED}$ could be measured with
about an hour of averaging.

\section{Systematic Effects}

\subsection{Drifting Birefringence}

One must check, though, for the sensitivity to systematic effects.  One of the
most problematic of these arises from the birefringence of mirrors.
Even a single bounce off of the high-quality dielectic optical coatings under
ideal conditions can introduce birefringence which drifts at 
$10^{-9}$\,s$^{-1}$ \cite{HaYeMa2000}. 
With respect to this systematic, our proposed high-frequency
lock-in measurement excels.  Both the probe and control pulses probe the same
mirror birefringence, only offset by about 10\,ns, so that the cumulative
differential spurious birefringence will be at the 
level of $10^{-13}$ and thus well below the vacuum birefrigence signal [for reference,
$10^{-9}$\,s$^{-1}$ $\times$ $10^{-8}$ s $\times \, (N=10,000) = 10^{-13}$].
In addition to drifting mirror birefringence, dynamic mirror birefringence associated with
cavity mirror motion should also be greatly mitigated since this noise is concentrated
at frequencies below $\sim1$~kHz~\cite{HaEtAl2019} and thus 
five orders of magnitude below our lock-in frequency.

\subsection{Thermal Noise}

There is good reason to believe that thermal noise in the optical cavity mirrors
could be limiting the PVLAS experiment~\cite{EjEtAl2020}.
However, there are good physical arguments for why such contributions should be
much smaller in our proposed experiment.

First, we consider thermal noise in the mirror substrates.  Length fluctuations 
due to Brownian motion should vary as $\nu^{-1/2}$ below a mechanical 
resonance, and as $\nu^{-5/2}$ above 
(see Sec.~2.2 of \cite{Go2000}).
Typical mirror mechanical resonances of the mirror substrate
are in the 10\,kHz range, and our lock-in frequency is $\sim$100\,MHz
(greatly above the mechanical resonances). Therefore, we would expect to
a suppression of such noise compared to the PVLAS 
experiment by roughly 10 orders of magnitude. 

Brownian noise in the optical coating (as opposed to in the substrate) is
theoretically predicted to have a  $\nu^{-1/2}$ behavior at high frequencies
and so would be reduced by a factor of about $\sqrt{10^8/10} \approx 3000$
compared with the PVLAS experiment (again, see \cite{EjEtAl2020}).  
However, as stated in Ref.~\cite{EjEtAl2020}, 
it is unclear why such such Brownian motion manifested
as birefringent noise (as opposed to simple interferometer length changes) in
the PVLAS experiment. Further, the PVLAS experiment observed a cutoff in the
supposed Brownian noise at 15\,Hz. If this poorly understood cutoff also exists
in our experiment, then the coating Brownian noise will be completely
negligible at 100 MHz. Therefore, we conclude that, while Brownian noise in
both the coating and substrate is not well studied in this context, it should
be highly suppressed in our proposed experiment.

The PVLAS experiment also provided evidence of
thermoelastic noise in the mirror coating~\cite{EjEtAl2020}. 
The thermoelestic noise density is expected to vary as $\nu^{-1/4}$ in the
low-frequency limit and as $\nu^{-1}$ in the high frequency limit \cite{BrVy2003}.  
The transition occurs when the frequency dependent diffusive heat transfer length
becomes smaller than the penetration depth into the optical coating 
($\sim1 \mu{}\mathrm{m}$), 
which occurs at a noise frequency of around 100 kHz. With this, the
thermoelastic noise should be around four orders of magnitude smaller at our
lock-in frequency of 100\,MHz when compared with the PVLAS experiment.  
We thus expect a large suppression of thermal noise 
in our apparatus.

\subsection{Stray Light}

Stray light from the pump field incident on the probe field detectors could imitate 
the LIVB signal since it would occur at the lock-in frequency.  
However, we note that 
there is no requirement that the probe and pump wavelengths are the same.  Therefore,
in addition to spatial filtering, 
spectral filtering can be used to greatly mitigate stray light 
at the polarimeter. A further obvious benefit of our apparatus, commensurate
with an idea initially formulated in Ref.~\cite{LuPe2004}, lies in the 
control of the overlap of the pump and probe pulses at the interaction volume.
Depending on the degree of overlap, the vacuum birefringence
sampled by the probe pulse can be modulated while keeping the thermal load on
the cavity mirrors nearly constant.  This can be used as an additional
confirmation that any observed signal is truly due to the pump versus probe
interaction.

\section{An Interferometric Measurement}

In order to maximize the expected $\chi_{\QED}$, the probe and control pulses
should enter the FEC at $45\degree$ to probe both $n_1$ and $n_2$. As shown in
Fig.~\ref{fig1}, there is a nonvanishing small angle of incidence on the second and 
third curved mirrors in the respective pump and probe cavities.
Due to the birefringence from the mirror at a non-zero angle of incidence, orthogonal
polarization components of the probe and control pulse will not, in general, be
simultaneously resonant within the cavity. One solution is to use only
two-mirror (as opposed to four-mirror) FECs, but this is technically
challenging (mirrors with 20\,cm diameters were envisioned 
in Ref.~\cite{LuPe2004}).
However, it is experimentally feasible to obtain degenerate polarization modes
by allowing one curved mirror to introduce a small deflection in a direction
perpendicular to $\hat{e}_1$ (in the $xz$ plane) while the other curved
mirror deflects the beam perpendicular to $\hat{e}_2$. For typical mirrors, a
$10\degree$ angle of incidence on the curved mirror introduces 
$\delta \phi \approx$ 10\,mrad of differential phase shift between the two orthogonal
polarization states.  Through tuning these angles carefully, suppression of the
static cavity birefringence should be possible. The frequency difference
between the two polarization modes $\Delta \nu$ can be related to the
differential phase shift, $\delta \phi$, through 
$\Delta \nu = \delta \phi/(2 \pi \tau)$ (see Ref.~\cite{LuPe2004}).
If the static birefringence can be suppressed by three orders-of-magnitude through 
a fine tuning of the mirror angles, then $\Delta \nu \lesssim 100$ Hz 
which is much smaller than 
the cavity resonance width of $\sim$3~kHz. 
This would create the requisite degeneracy between the 
polarization states.

However, this tuning could be experimentally challenging.  When the mirrors are
subjected to a high fluence, it is likely that the optimal angle on the mirrors
would change slightly due to the thermal load on the mirrors~\cite{HaYeMa2000}.
Therefore, as is evident from Fig.~\ref{fig1}, a preferred approach to
eliminate the challenge of non-degenerate polarization modes, could be to only
couple one polarization state into the cavity and perform interferometric
measurements of $n_1$ and $n_2$ separately~\cite{Ah2023,Ah2018}.  While one
might assume that a direct measurement of the birefringence ({\em i.e.}, 
$n_1 - n_2$) is highly preferred due to its differential nature, in our scheme, the 
differential comparison of the highest fidelity 
is between the probe and control pulses and
not between the different polarization states.  Therefore, it is also
reasonable to measure $n_1$ and $n_2$ (given
in Eqs.~\eqref{n1exp} and~\eqref{n2exp}) individually, 
with the additional advantage of refining the QED test 
inherent to our proposal. In this case, probe-laser
light polarized either in the $\hat e_1$ or the $\hat e_2$ directions (and not
a superposition) is sent to the probe cavity. In addition, the incident 
probe-pulse train (sampled by the first beam splitter encountered after the probe
laser) is interferred with the leakage pulse train from the enhancement cavity
(see Fig.~\ref{fig1}).  The probe and control pulses leaking from the cavity
must be correctly timed to interfere with the respective sampled pulses and
phase-locked, which requires the active adjustment of a delay line.  This would 
ensure that the power at the output port 
of the interferometer remains approximately at a 
defined nominal value.  However, the active control would
only respond at the low frequencies of typical lab vibrations ($< 1$ kHz) and 
would therefore, have no effect on the signal.

%
%
\section{Conclusions}

In summary, we have shown that, by using 
interferometric techniques involving pump, probe and control
ultrashort laser pulses of repetition rate $\sim$100\,MHz, which are 
cavity-enhanced to $\sim$$1$\,mJ pulse energies,
the high-frequency detection of vacuum birefringence
appears feasible. Our proposal profits from the availability 
of strong instantaneous magnetic fields in the pump laser field,
which surpass achievable static field strengths by 
orders of magnitude.  The use of high-repetition-rate FECs 
allow for a {\it cumulative} polarization rotation observed on the probe laser, 
which is absent in petawatt laser experiments. The two 
decisive advantages of our proposal are {\em (i)} the high repetition rate 
which allows for the mitigation of technical noise 
through a high-frequency differential measurement,
and {\em (ii)} the possibility of measuring the vacuum
birefringence for individual polarization components,
{\em i.e.}, in the sense of Eq.~\eqref{n1n2exp}, 
the possibility of measuring not only 
$\Delta n_2 - \Delta n_1$, but also, $\Delta n_2/\Delta n_1 = 7/4$.

\section*{Acknowledgements}

We would like to thank 
Antonino di Piazza, Siu Au Lee, Christian Sanner, 
Sam Brewer, Jacob Roberts, Jun Ye and 
Thomas Allison for useful 
conversations and comments on the manuscript. 
This research has been supported 
by the National Science Foundation
(Grants PHY--2110294 and PHY--2207298) 
and by the Gordon and Betty Moore Foundation, 
GBMF12955, Grant DOI: 10.37807/GBMF12955.

\appendix

\section{Shot-Noise Limited Measurements}
\label{appa}

Let us now consider the pulse after the 
interaction, i.e., with a phase difference 
$\delta$ between the different polarization 
components and an eccentricity angle $\chi_\QED$.
We consider a pulse with Jones vector \cite{Co2005} given by
$\frac{1}{\sqrt{2}}
\begin{pmatrix}
1 \\
\ee^{\ii \delta} 
\end{pmatrix}$
incident on a polarimetry setup consisting of a 
quarter waveplate with its fast axis at an angle $\phi$ with respect to the $y$-axis,
followed by a polarizer set at $-45^\circ$.  As before, $\delta$ is the 
phase to be measured arising 
from the vacuum birefringence, which is assumed to be small. In this case, it 
is straightforward to show that the number of detected 
photons is given by
\begin{equation}
N_d=\frac{\cos^2{(2 \phi)}+\cos{(2 \phi)} \, \delta}{2} \, N_\gamma X,
\end{equation}
where $N_\gamma$ is the total number of photons in the pulse, and $X$ is the quantum
efficiency of the detector.  In our differential measurement, the signal is
the difference in the output for probe and control pulses (where 
$\delta$ for the control pulse is set to zero), which is given by
\begin{equation}
\mathrm{Signal}=\frac{\cos{(2 \phi)}}{2} \delta N_\gamma X,
\end{equation}
while the noise is the square root of the total number 
of detected photons, given by
\begin{equation}
\mathrm{Noise}=\cos{(2 \phi)} \sqrt{N_\gamma X}.
\end{equation}
From this, we see that the signal-to-noise ratio is given by
\begin{equation}
\mathrm{SNR}=\frac{\sqrt{N_\gamma X} \delta}{2}=\sqrt{N_\gamma X} \chi_{\QED}.
\end{equation}
While the signal-to-noise calculated from shot noise is independent of $\phi$,
the adjustment of $\phi$ allows one to set the intensity on the photodetector
appropriately (i.e. avoiding satuation/damage of the detector while also
mitigating the effects of dark counts).

The calculation proceeds in an analogous fashion for the interferometric
determination of $\delta$. In that case,
the function of $\phi$ is performed by the phase offset in the interferometer,
which can also be appropriately chosen to avoid detector saturation.

\section{Expected Signal with Gaussian Pulses}
\label{appb}

Our above estimate for 
$\chi_\QED$ approximates the ellipticity
in the proposed experiment. This value is moderately reduced 
when taking into account the Gaussian spatiotemporal nature of the 
colliding optical pulses. This is because a significant fraction of the energy 
in the probe pulse does not sample the peak $\Delta n_{\QED}$ produced by the 
pump pulse. It is thus desirable to provide a more accurate
estimate of the expected signal of 
vacuum birefringence within approximations which still allow one to 
obtain an analytic formula, to model the expected signal in the 
experiment. 
Let $P$ be the power of a Gaussian pulse,
with the radial intensity profile
\begin{equation}
I(r) = \frac{2 P}{\pi w^2} \, \exp\left( - \frac{2 r^2}{w^2} \right) \,.
\end{equation}
Here, $w$ is the waist parameter of the Gaussian beam.
One can easily show that 
$P = 2 \pi \int_0^\infty \dd r \, r \, I(r)$,
and the power $P$ is recovered as the radial integral 
over the intensity over a cross-sectional area.
We assume the waist to depend on the $z$ coordinate as follows,
\begin{equation}
w = w_z = w_0 \, \sqrt{ 1 + \left( \frac{z}{z_\RR} \right)^2 } \,,
\quad
z_\RR = \frac{\pi w_0^2}{\lambda} \,,
\end{equation}
where $z_\RR$ is the Rayleigh range.
In order to model the intensity profile of the 
pump pulse (traveling in the positive $z$ direction), 
we parameterize the time duration of the Gaussian pulse
by a parameter $t_0$,
\begin{equation}
I_{\rm pump}(\vec r, z, t) =
\frac{2 \calE_{\rm pump} }{\pi w_z^2} \,
\frac{\ee^{- (t - z/c)^2 / t_0^2}}{\sqrt{\pi} t_0} \,
\exp\left( -\frac{2 r^2}{w_z^2} \right) \,.
\end{equation}
The parameter $t_0$, for both probe and pump,
is assumed to be roughly equal to 
$250 \times 10^{-15} \, {\rm s}/(2 \sqrt{\ln(2)})$,
corresponding to a full width at half maximum of 
$250$ femtoseconds.
The energy of the pump pulse is denoted
as $\calE_{\rm pump}$.
The averaged (over a laser oscillation cycle)
squared electric field of the pump laser
is given as
\begin{equation}
\left< \vec E^2_{\rm pump} \right> =
\frac{ I_{\rm pump}(\vec r, z, t)}{\epsilon_0 \, c} \,,
\end{equation}
resulting in an averaged (over a pump laser oscillation cycle)
vacuum birefringence of
\begin{equation}
\Delta n_{\rm QED}(\vec r, z, t) \approx
\frac{\alpha}{30\pi}
\frac{4 \left< \vec E^2_{\rm pump} \right> }{E^2_{\mathrm{cr}}}
= \frac{\alpha}{30\pi}
\frac{ 4 I_{\rm pump}(\vec r, z, t)}%
{\epsilon_0 \, c \, E^2_{\mathrm{cr}}} \,,
\end{equation}
where we appeal to Eq.~\eqref{nQED},
within the approximation of vanishing $\theta$.

For simplicity, in order for our final 
result to be expressible in a compact closed 
analytic form, we assume the waist and the Rayleigh range
of the probe pulse 
to be the same as that of the pump beam,
\begin{equation}
I_{\rm probe}(\vec r, z, t) = 
\frac{2 \calE_{\rm probe} }{\pi w_z^2} \,
\frac{\ee^{- (t + z/c)^2 / t_0^2}}{\sqrt{\pi} t_0} \,
\exp\left( -\frac{2 r^2}{w_z^2} \right) \,.
\end{equation}
Here, $\vec r = x \hate_x + y \, \hate_y$ 
is the coordinate in the plane perpendicular to 
the propagation direction $z$.
The probe pulse is assumed to be traveling in the negative $z$ direction,
and the probe pulse energy is $\calE_{\rm probe}$.
The current density of the probe photon 
is obtained as follows,
\begin{equation}
j_{\rm probe}(\vec r, z, t) = 
\frac{I_{\rm probe}(\vec r, z, t)}{\calE_{\rm probe}} \,,
\end{equation}
which is independent of the probe-field intensity. For the analysis of colliding femtosecond pulses,
$\chi_\QED$ as given in Eq.~\eqref{chiQED}
needs to be brought into differential form,
$\chi_{\QED} = \int \dd \chi_{\QED}$, where
\begin{multline}
\chi_{\QED} = 
\int \dd L \, \dd A \; \dd t \; j_{\rm probe} \; \Delta n_{\QED} 
= \frac{N \pi}{\lambda} \times
\\
\int\limits_{-\infty}^\infty \dd z 
\int\limits_0^\infty \dd r \, (2 \pi r) \,
\int\limits_{-\infty}^\infty \dd t \;
j_{\rm probe}(\vec r, z, t) \;
\Delta n_{\rm QED}(\vec r, z, t) \,.
\end{multline}
Here, $\dd A = \dd r (2 \pi r)$ is the cross-sectional area.
The final result for the ellipticity $\chi_{\rm QED}$ is 
\begin{subequations}
\label{chi}
\begin{align}
\chi_{\QED} =& \; \frac{\alpha \, N \, \calE_{\rm pump} \sqrt{2 \pi} }%
{15 c \, \epsilon_0 \, t_0 \, \lambda^2 \, E^2_{\mathrm{cr}} } \,
\exp( X^2 ) \, \mathrm{erfc}( X ) \,,
\\
X =& \; \frac{\sqrt{2} \pi w_0^2}{c \, t_0 \, \lambda} \,,
\end{align}
\end{subequations}
where $\mathrm{erfc}$ is the complementary error function~\cite{AbSt1972}.
It is straightforward to convince oneself that 
different probe and pulse durations can easily be included with the simple replacement
$t_0 \rightarrow \sqrt{t_0^2/2+t_0'^2/2}$, where $t_0$ 
remains the pulse duration of the probe, 
and $t_0'$ is the pulse duration of the pump.
With the parameters $\alpha \approx 1/137.036$,
$\calE_{\rm pump} = 0.0016 \, {\rm J}$,
$c = 3 \times 10^{-8} \, {\rm m}/{\rm s}$,
$\epsilon_0 = 8.854 \times 10^{-12} {\rm F}/{\rm m}$,
$\lambda = 1.04 \times 10^{-6} \, {\rm m}$,
$w_0 = 4.0 \times 10^{-6} \, {\rm m}$,
$t_0 = t'_0 = 250 \times 10^{-15} {\rm s}/(2 \sqrt{\ln(2)})$,
$N = 10,000$, and
$E_{\mathrm{cr}} = 1.3233 \times 10^{18} \, {\rm V}/{\rm m}$,
one obtains $\chi_{\QED} = 8.23 \times 10^{-12}$. The result~\eqref{chi} is manifestly independent 
of the relative phase of the pump and probe laser pulses
in the interaction region, and depends 
only on the time durations $t_0$ (and conceivably
$t'_0$) of the pump and probe pulses. 
Let us try to understand this 
result intuitively. The general formulas
for the laser-induced vacuum birefringence
given in Eqs.~\eqref{n1exp} and~\eqref{n2exp}
depend on the instantaneous $\vec E$ and $\vec B$ fields 
of the pump laser, and, in that sense,
are phase-dependent. As is well known,
the nodes and maxima of the electric and magnetic
fields of traveling electromagnetic waves
are aligned. However, in the overlap region
of the probe and pump pulses, we may average
over a laser oscillation period, and the signal is 
given by an integration over the photon-number
current density of the probe pulse,
multiplied by the cycle-averaged vacuum birefringence
due to the pump pulse. For pump pulses
of $1.19$\,eV and durations of about 250\,fs, 
with about 70 laser oscillations,
this approximation is well justified and 
leads to the manifestly phase-independent result
given in Eq.~\eqref{chi}.

\pagebreak

\end{document}